# A Language for Autonomous Vehicles Testing Oracles

Ana Nora Evans
AnaNEvans@virginia.edu
University of Virginia

Mary Lou Soffa
soffa@virginia.edu
University of Virginia

Sebastian Elbaum
selbaum@virginia.edu
University of Virginia

## Abstract

Testing autonomous vehicles (AVs) requires complex oracles to determine if the AVs behavior conforms with specifications and humans' expectations. Available open source oracles are tightly embedded in the AV simulation software and are developed and implemented in an ad hoc way. We propose a domain specific language that enables defining oracles independent of the AV solutions and the simulator. A testing analyst can encode safety, liveness, timeliness and temporal properties in our language. To show the expressiveness of our language we implement three different types of available oracles. We find that the same AV solutions may be ranked significantly differently across existing oracles, thus existing oracles do not evaluate AVs in a consistent manner.

## 1 Problem and Motivation

Oracles are predicates that determine if a program is correct with respect to a specification [4]. In the AV domain, the program is replaced by the AV's behavior, since the software's output is a a trace of timed series of actuation commands for brakes, throttle and steering. Furthermore, a numeric output of the oracle can be used to rank AV solutions. Thus, a *scoring oracle* is a *program* that calculates a real value score from an execution trace and relevant information about the environment [5].

We propose a domain specific language, called an *Oracle Definition Language* (ODL), which facilitates the definition of AV oracles, independent of the simulators and AV solutions and that is expressive enough to encode available open source oracles. An AV oracle can be expressed in ODL as a program, called *oracle definition* (OD) which processes an execution trace either generated in a simulated environment or collected on a real world drive, and outputs a score.

To design the ODL, we studied available open source oracles, including ones from education [6], research [2], and industry [1, 3]; most commercial oracles are not broadly available. We identified four types of properties that the language must support: (1) safety (something bad does not happen, *e.g.*, the speed is always lower than the speed limit), (2) liveness (something good happens, *e.g.*, reaching the destination), (3) timeliness (something is bad only if it lasts longer than a time interval, *e.g.*, the vehicle drives on a line for more than three seconds), (4) temporal properties (ordered sequences of properties, *e.g.*, the AV decelerates for at least two seconds before collision).

The ODL is designed to facilitate testing and analysis techniques to increase confidence in the oracle's correctness with respect to an oracle specification and methods to generate realistic execution traces to improve the test suite quality.

## 2 Approach

In our method, depicted in Figure 1, the testing analyst uses the domain specific language, ODL, to define an oracle program (OD). The *Trace* and OD are inputs to the *Scoring Oracle* (an ODL interpreter) that assigns a real valued *Score* to the given inputs. In particular, the OD program is a collection of *scoring functions*, each defining a property of the input trace and the corresponding changes to the function's score to be made when the property is satisfied.

Next, we introduce the definitions of the ODL concepts.

**Definition:** A *scoring function* is a function $f : Traces \to \mathbb{R}$ that associates a real number to a trace. For example, a function may compute a score by counting the number of collisions appearing in a trace multiplied by a negative five.

**Definition:** A *summarizing function* is a function $f : \mathbb{R}^k \to \mathbb{R}$, that associates a real number to a set of scores, where $k$ is the number of scoring functions being summarized. For example, a summarizing function may add the result of all scoring functions.

**Definition:** An *oracle definition* (OD) consists of a set of scoring functions and a summarizing function.

**Definition:** A *scoring oracle* (SO) is a function $f : (OD, tr) \to \mathbb{R}$, associating an oracle definition (OD) and a trace with a score. For example, given the trace $tr = [[1, 0, t], [2, 0, t]]$ where the last element in each tuple is the presence of a collision, an OD with a single scoring function that counts the number of collisions and multiplies it by 5, the SO will produce a value of 10.

The ODL introduces a set of useful abstractions for the testing analyst. The input trace format hides the complexities of extracting information from the simulator. The summarizing scoring function can be used to define various methods of calculating the score from the score given by the properties checked. The scoring function abstraction can be used to identify parts of the trace satisfying a property, calculate a value dependent on the part the trace, and aggregate the values into a score using predefined functions.

We illustrate the expressiveness of the ODL with four examples, one for each type of property.[1]

**Safety Property:** The `speeding` scoring function in Listing 1 subtracts one for each trace element with speed higher than the speed limit (`MAX_SPEED`). The `event` is a Boolean expression. The variable `speed` holds the value of the corresponding field of the trace message. The `frequency` defines how the score is updated with the value of the evaluation of the arithmetic expression `action`

```
1 speeding = scoring_function( event = speed > MAX_SPEED,
2     action = -1, frequency = action_sum)
```

---

[1]We also developed a grammar and a high level operational semantics.

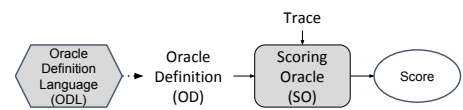

**Figure 1: General Framework.**

<div align="center">Listing 1: Safety Property.</div>

**Liveness Property:** The `arrival_test` scoring function in Listing 2 has the score one if the AV is within twelve meters of the `destination`. The `position` variable contains the Cartesian map coordinates of the AV. The `distance` function calculates the Euclidean distance between two points on the map. The value `first` of the `frequency` means that the score is updated only the first time the event is true.

```
1 arrival_test = scoring_function(
2     event = distance(position, DESTINATION) < 12,
3     action = 1.0, frequency = first)
```

<div align="center">Listing 2: Liveness Property.</div>

**Timeliness Property:** The `lane_keep` scoring function in Listing 3 subtracts one for each maximal sequence in which the AV drives on a line for more than three seconds. The new parameter, `condition` changes the semantics of score update as follows: a maximal sequence of messages for which the `event` is true is created and the score is updated only if this sequence satisfies the Boolean expression `condition`. In Listing 3, the condition is the length in time of the maximal sequence is greater than three seconds.

```
1 lane_keep = scoring_function(
2     event = (road_normal > LW-TH and road_normal < LW+TH) or
3         (road_normal > 2*LW-TH and road_normal < 2*LW+TH),
4     condition=seq_time > 3, action=-1, frequency=action_sum)
```

<div align="center">Listing 3: Timeliness Property.</div>

**Temporal Property:** The `collisions` scoring function in Listing 4 updates the score only if the scoring function `deceleration` was activated within half of second of the collision event. The ODL uses an Observer pattern to implement temporal properties. A `notification` is sent when the `condition` of the `deceleration` scoring function is satisfied and the value of the internal variable `expiration` of the `collisions` scoring function is set to half of second. This value is decreased every time a message is processed and the `condition` of `deceleration` is satisfied only when the `expiration` is still positive.

```
1 collisions = scoring_function(
2     event = collision and expiration > 0,
3     action = 1.0, frequency = all_sum)
4 deceleration = scoring_function(
5     event = acceleration < 0 and not collision,
6     condition = seq_time > 2, frequency = all_sum,
7     notifications = [(collisions, [(expiration, 0.5)])])
```

<div align="center">Listing 4: Temporal Property.</div>

## 3 Results and Contributions

To evaluate our design we modified the the simulator for the path planning project of the Udacity's Self Driving Car Engineer nanodegree to generate execution traces. The AV solutions are students' solutions to the path planning project (474 open source projects). The ODs are based on the following AV oracles: (OD1) the grading rubric for the path planning project [6]; (OD2) the Carla Challenge at CVPR 2019 [2]; (OD3) testing oracles from the Apollo (Baidu) open source framework for AV solutions development [1].

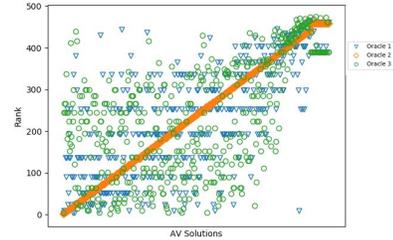

**Figure 2: Rankings given by ODs.**

The ODL is implemented as a domain specific language embedded in Python. For each OD, we calculate and average score over thirty executions, and use it to assign a rank to each AV solution. Next, we use the Spearman rank-order coefficient to understand the statistical dependence between the ranks in pairs of ODs. We found that one pair has a strong correlation, while the other two pairs have only a moderate one. Figure 2 offers a visual representation of the Spearman coefficient. The x axis represents AV solutions ordered by the rank of OD2, the one that was originally designed to evaluate AV's in a competition. The y axis is the rank from zero to 474, with zero the best ranking.

Our main contribution is a domain specific language, the ODL, for the definition of AV oracles in a unified manner, independent of the simulator and AV solution. The language is expressive enough to encode safety, liveness, timeliness and temporal properties. Our evaluation demonstrates that available AV oracles rank solutions differently enough to support the need for testing and analysis tools.

## References

[1] Baidu Apollo team. 2017. Apollo: Open Source Autonomous Driving. https://github.com/ApolloAuto/apollo. Accessed: 2020-01-20.

[2] Carla. 2019. Carla Challenge at CVPR 2019. https://carlachallenge.org/.

[3] A. Censi, K. Slutsky, T. Wongpiromsarn, D. Yershov, S. Pendleton, J. Fu, and E. Frazzoli. 2019. Liability, Ethics, and Culture-Aware Behavior Specification using Rulebooks. In *2019 International Conference on Robotics and Automation (ICRA)*. 8536–8542. https://doi.org/10.1109/ICRA.2019.8794364

[4] W. E. Howden. 1978. Theoretical and Empirical Studies of Program Testing. *IEEE Transactions on Software Engineering* SE-4, 4 (July 1978), 293–298. https://doi.org/10.1109/TSE.1978.231514

[5] Claudio Menghi, Shiva Nejati, Khouloud Gaaloul, and Lionel C. Briand. 2019. Generating Automated and Online Test Oracles for Simulink Models with Continuous and Uncertain Behaviors. In *Proceedings of the 2019 27th ACM Joint Meeting on European Software Engineering Conference and Symposium on the Foundations of Software Engineering (ESEC/FSE 2019)*. Association for Computing Machinery, New York, NY, USA, 27–38. https://doi.org/10.1145/3338906.3338920

[6] Udacity. 2019. CarND-Path-Planning-Project. https://github.com/udacity/CarND-Path-Planning-Project.